\documentstyle[prb,aps,preprint,epsf]{revtex}

\begin{document}

\title{Auger Effect in the High-Resolution Ce 3d-edge
Resonant Photoemission}

\author{E.-J. Cho}
\address{Department of Physics, Chonnam National University, Kwangju 500-757,
Korea}

\author{R.-J. Jung, B.-H. Choi, and S.-J. Oh}
\address{School of Physics and Center for Strongly Correlated Materials Research\\
Seoul National University, Seoul 151-742, Korea }

\author{T. Iwasaki, A. Sekiyama, S. Imada, and S. Suga}
\address{Department of Material Physics, Graduate School of Engineering Science, Osaka
University, Toyonaka, Osaka 560-8531, Japan}

\author{T. Muro}
\address{Japan Synchrotron Radiation Research Institute, Sayogun,
Hyogo 679-5198, Japan}

\author{J.-G. Park, and Y.S. Kwon}
\address{Department of Physics, SungKyunKwan university, Suwon 440-746, Korea}

\date{Received \hspace*{30mm}}

\maketitle

\begin{abstract}

The bulk-sensitive Ce 4$f$ spectral weights of various Ce
compounds including CeFe$_2$, CeNi$_2$, and CeSi$_2$ were obtained
with the resonant photoemission technique at the Ce $3d$-edge.  We
found the lineshapes change significantly with the small change of
the incident photon energy.  Detailed analysis showed that this
phenomenon results primarily from the Auger transition between
different multiplet states of the Ce $\underline{3d_{5/2}}4f^2$
(bar denotes a hole) electronic configuration in the intermediate
state of the resonant process.  This tells us that extra care
should be taken for the choice of the resonant photon energy when
extracting Ce 4$f$ spectral weights from the Ce 3$d$-edge resonant
photoemission spectra.  The absorption energy corresponding to the
lowest multiplet structure of the Ce $\underline{3d_{5/2}}4f^2$
configuration seems to be the logical choice.

\end{abstract}
\pacs{PACS numbers: 71.28.+d, 73.20.At, 79.60.-i}


\section{INTRODUCTION}

Investigations on the electronic structures of strongly correlated
materials such as $d$ and $f$ electron systems have been very
active lately.\cite{Lawrence81,Stewart84} One of the perhaps
oldest problems in this category is the origin of the $\alpha
\Leftrightarrow \gamma$ phase transition of Ce metal and its
compounds, which has been studied for more than 50 years but still
remains controversial.\cite{Pauling,Allen86,Malterre96}
Photoemission spectroscopy, among various experimental techniques,
is a very powerful tool that can directly probe the electronic
structures of solids. Indeed in the case of Ce problem as well,
the study of 4$f$ spectral weights of various Ce compounds using
resonant photoelectron spectroscopy (RPES) technique at the Ce
4$d$-edge contributed decisively to the understanding of this
famous phase transition.  This technique is necessary to enhance
the 4$f$ electron emissions and separate them from contributions
of other conduction electrons,\cite{Allen81} and utilizes the
following process

\begin{eqnarray*}
 4d^{10}4f^1 +  \hbar \omega \ \rightarrow  \ 4d^94f^2  \ \rightarrow  \ 4d^{10}4f^0 +
 photoelectron.
\end{eqnarray*}

The conventional wisdom that came out of these studies, at least
until early 1990's, was as follows; The major factor
distinguishing between $\alpha$ and $\gamma$ phases of Ce and its
compounds is the strength of the hybridization between Ce 4$f$
level and the conduction bands, and the disappearance of the local
magnetic moment in the $\alpha$ phase results from the Kondo
phenomenon of the singlet formation due to the hybridization of a
localized 4$f$ electron with conduction electrons, and not from
the promotion of the 4$f$ electron to the empty state above the
Fermi level.\cite{Pauling,Allen86,Malterre96}

After the high resolution RPES was developed in the early 1990's,
however, some groups questioned this conventional wisdom and
proposed a new interpretation of the 4$f$ spectral weights of Ce
compounds.\cite{Joyce92,Blyth93} The main point of those groups
was that the temperature or the material dependence of the 4$f$
spectral weights of Ce compounds near the Fermi level, which had
been interpreted as the tail of the Kondo
resonance,\cite{Gunnarsson83} did not follow the Kondo scenario
which stipulates that the Kondo temperature is the universal
controlling parameter that determines the strength of the Kondo
resonance near the Fermi level in different materials or at
different measuring temperatures.  This controversy generated many
more careful studies on the 4$f$ spectral weights of various Ce
compounds by several research groups,
\cite{Malterre92,Grioni97,Garnier97,Yang96,Kim97} some utilizing
RPES technique and others using ordinary ultraviolet photoemission
(UPS) technique with He$_I$ (h$\nu$ = 21.2 eV) and He$_{II}$
(h$\nu$ = 40.8 eV) sources.

Most of these later studies also supported Kondo resonance
scenario, but one important issue which has not been settled so
far is the surface effect and how to separate the surface from the
bulk contributions.\cite{Duo}  Since the electron kinetic energy
at the Ce $4d$-edge RPES is about 120eV, and the electron mean
free path at this kinetic energy is only about
5\AA,\cite{Seah79,Wagner80} the surface contribution in the 4$f$
spectral weights obtained with $4d$-edge RPES technique is
expected to be very much significant, even larger than the bulk in
some cases.  The situation with the ordinary UPS data with He$_I$
(h$\nu$ = 21.2 eV) and He$_{II}$ (h$\nu$ = 40.8 eV) sources is not
much different. Hence to compare quantitatively the experimental
photoemission spectra with those expected from the Kondo
temperature of bulk materials, it is imperative to be able to
extract the bulk contributions from the experimental data. Many
approaches have been tried with the Ce $4d$-edge RPES or ordinary
UPS data,\cite{Yang96,Kim97} but the results were not conclusive
enough since each method has its own drawbacks and uncertainties.

Recently a powerful new method to extract reliable bulk Ce 4$f$
spectral weights has become possible thanks to the advent of the
new beamline at SPring-8, Japan.\cite{Saitoh} It has been
well-known that we can reduce the surface contribution in
photoemission spectra by using high energy photon source. For
example, if around 880eV photon source is used in RPES, the
electron mean free path of the photoelectrons from Ce 4$f$ levels
will be about 20 \AA.\cite{Seah79,Wagner80} Since those 880eV
photon energies can excite a Ce $3d$ core-electron to an empty
4$f$ level (Ce 3$d$ absorption edge), the 4$f$ spectral weight is
also enhanced due to the resonance phenomenon as in the case of Ce
4$d$-edge RPES.  Hence we can obtain much more bulk-sensitive Ce
4$f$ spectral weights by using those high photon energies in this
so-called Ce 3$d$-edge RPES technique. In fact, this technique had
been already tried more than 10 years ago,\cite{Laubschat90} but
the energy resolution in the Ce 3$d$-edge RPES at that time was
not good enough to give conclusive results.  Recently, Sekiyama
{\it et al.} succeeded to obtain the total energy resolution
including energy analyser at the Ce $3d$ edge photon source less
than 100meV with high photon flux in the undulator beamline of
SPring-8 and study high-resolution bulk-sensitive Ce spectral
weights in CeRu$_2$.\cite{Sekiyama00} This opened up a new
possibility to extract reliable bulk-sensitive spectral weights in
many interesting Ce compounds, and several papers have already
appeared utilizing this Ce 3$d$-edge RPES
technique.\cite{Sekiyama00,Yang00,Jung01}

However, in the course of this study, we realized that the
experimental photoemission data depends sensitively on the exact
incident photon energy near the Ce 3$d$-edge, and the difference
of spectral lineshape is significant enough to affect data
interpretation. Also the new Ce 3$d$-edge RPES data were found to
be not completely consistent with the old data taken in early
1990's, even after the difference of the energy resolutions is
taken into account.  To derive reliable conclusions from the Ce
3$d$-edge RPES data, therefore, it is necessary to understand the
origin of these differences and changes.  This paper is an attempt
to understand the origin of these phenomena by systematically
studying Ce 3$d$-edge RPES of several Ce compounds. We will show
in this paper data on CeFe$_2$, CeNi$_2$, and CeSi$_2$ only,
although we studied many more Ce compounds which gave fully
consistent results with the conclusion of this paper. The three Ce
compounds to be discussed here have various Kondo temperature, the
universal parameter which controls many physical properties of Ce
compounds in the Kondo resonance scenario. The Kondo temperatures
of intermetallic compounds CeFe$_2$ and CeNi$_2$ are larger than
about 500 K, and that of CeSi$_2$ compound is about 45
K.\cite{Allen86,Malterre96} This paper will mainly focus on the
general phenomena of the Ce 3$d$-edge RPES, and the electronic
structures of individual compounds will be discussed in separate
publications.\cite{Jung,Choi}

\section{EXPERIMENTAL}

Polycrystalline samples of CeFe$_2$, CeNi$_2$, and CeSi$_2$ were
made by the arc melting method under Ar gas environment with
constituent elements Ce, Fe, Ni, and Si whose purity was better
than 99.9 \%. The homogeneity of samples was checked after
annealing with x-ray diffraction.

Resonant photoemission spectra of the valence bands at various
photon energies near the Ce $3d$-edge (h$\nu \sim$ 880 eV) along
with the Ce $3d$-edge x-ray absorption spectrum (XAS) were
obtained at the beamline BL25SU in the SPring-8 of
Japan.\cite{Saitoh} The base pressure of an analysis chamber was
about 3 $\times$ 10$^{-10}$ mbar or better. The samples were
scraped with a diamond file to get clean surfaces, and every time
the cleanness of sample surface was checked with the O $1s$
core-level spectrum. If the oxygen contamination peak was found
noticeable in O $1s$ core-level spectrum, the samples were scraped
again with the diamond file.  All spectra were measured at the
temperature of 20 K, and the scraping was also done at 20 K.  The
XAS measurements were done by the total electron yield mode, and
the energy resolution was about 80 meV.  Scienta 200 analyzer was
used for the electron energy analysis for PES measurements, and
the total resolution of the PES spectra was better than 100 meV.

\section{DATA AND DISCUSSIONS}

Fig. 1 shows the Ce M$_V$ XAS data representing $3d_{5/2}
\rightarrow 4f $ transition for CeFe$_2$, CeNi$_2$, and CeSi$_2$
compounds.  The $3d_{5/2}$ XAS of Fig. 1 consists of the main peak
structure near $h\nu $ = 882 eV, and the satellite peak near $h\nu
$ = 887 eV.  The main peak structure comes from $"3d^{10}4f^1"
\rightarrow "3d^94f^2" $ transition, and the satellite peak
$"3d^{10}4f^0" \rightarrow "3d^9f^1" $ transition.\cite{Fuggle83}
The mark of " " means approximate interpretation, since the
complete eigenstate of each peak is composed of many body
configurations. The lineshape of the main peak near 882 eV is
primarily determined by the multiplet structures of the $
\underline{3d}4f^2 $ electronic configuration, where the underline
represents a hole. At the bottom of the figure is shown the
theoretically expected XAS line spectrum, which is calculated by
considering all the multiplets of the $\underline{3d_{5/2}}4f^2$
electronic configuration along with the appropriate transition
probability from the ground state.\cite{Jo88}  We can see that the
theoretical calculation compares favorably with the experimental
XAS data, especially for the case of CeSi$_2$, although we can
also notice that the details of the main peak lineshape and the
satellite peak intensity near 887eV change somewhat depending on
the compounds. We will deal with this phenomenon and its
implication on the electronic structures systematically in a
separate publication.\cite{Choi} In this figure, the symbols A to
G designate the incident photon energies used to obtain the $4f$
spectral weights with the Ce $3d$-edge RPES to be shown in
subsequent figures.

Fig. 2 represents RPES spectra near the Fermi level of CeFe$_2$
obtained with various incident photon energies corresponding to
the symbols from A to G in Fig. 1.  All through these photon
energies the Ce 4$f$ spectral weights are resonantly enhanced by
the following process.
\begin{eqnarray*}
 3d^{10}4f^1 +  \hbar \omega \ \rightarrow \  3d^94f^2 \  \rightarrow  \ 3d^{10}4f^0 +
 photoelectron.
\end{eqnarray*}
Since the contribution from the Ce 4$f$ emissions is at least ten
times larger than those from other conduction electrons at these
resonance energies, we can safely neglect the other conduction
electron contributions and consider the spectra of fig. 2 as the
Ce 4$f$ spectral weights near the Fermi level, as in the case of
Ce 4$d$-edge RPES.\cite{Allen86,Allen81} However, the bulk
contribution of Ce 4$f$ spectral weights of Fig. 2 is much larger
than that of $4d$ edge RPES, since the electron mean free path at
$3d \rightarrow 4f$ transition energies is much
longer.\cite{Seah79,Wagner80}

The spectra in Fig. 2 consist of a peak near the Fermi level and a
shoulder at about 2eV.  The shoulder near 2eV comes from the $f^1
\rightarrow f^0$ transition, whereas  the peak near the Fermi
level comes from $f^1 \rightarrow f^0\underline{c}$
transition(\underline{c} means the hole in the conduction band)
which corresponds to the tail of the Kondo resonance
peak.\cite{Allen86,Malterre96,Joyce92,Blyth93,Gunnarsson83,Malterre92,Grioni97,Garnier97,Laubschat90,Sekiyama00,Yang00,Jung01}
The detailed analysis on those peaks and their relation to the
electronic structures of CeFe$_2$ have been dealt with in other
papers.\cite{Yang00,Jung01}  Here we focus on the change of the
4$f$ spectral lineshapes depending on the incident photon energy.
We clearly see that as the photon energy is increased from A
(881.76eV) to F (882.48eV), which is only 0.72eV apart, the width
of the peak near the Fermi level is significantly broadened. To
see these changes of spectral weights more quantitatively, we
subtracted the spectrum at A from data taken at higher photon
energies after normalizing each spectrum at its maximum intensity.
The resulting difference spectra at various photon energies are
shown in Fig.3.

We find two noticeable features in every difference spectrum of
Fig.3 --- a broad peak below the Fermi level E$_F$ and a sharp dip
at E$_F$.  The maximum of the broad peak moves to the high binding
energy side as the incident photon energy is increased. In fact,
we note that the binding energy of the maximum position is exactly
the same as the difference of two incident energies, which is
indicated by the arrows. This tells us the kinetic energy of the
broad peak remains the same regardless of the incident photon
energy, a telltale signature of Auger emission.  Hence, we can
conclude that this broad peak results from the Auger transition,
and that the apparent change of the spectral lineshapes depending
on the incident photon energy is due to these Auger emissions
overlapping the resonant 4$f$ photoemissions. This Auger process
must be the Coster-Kronig type between different multiplets of the
$3d^94f^2$ configuration in the intermediate state of RPES.

To obtain genuine Ce 4$f$ spectral weights from RPES data,
therefore, it is necessary to avoid the contribution from these
overlapping Auger emissions.  The best way to achieve this is
probably to take the spectrum at the incident photon energy
corresponding to the lowest multiplet of the intermediate state.
Indeed, we can see in the figure that the difference spectrum
between B and A is nearly flat, implying that the spectral shape
is nearly identical for data taken near the lowest multiplet.  We
have also confirmed that the spectra taken at incident photon
energies slightly below A remain identical in lineshapes as that
of A.   This fact gives us an important lesson in extracting
bulk-sensitive Ce 4$f$ spectral weights using the 3$d$-edge RPES
technique --- that is, it is dangerous to use the incident photon
energy which simply gives maximum resonance effect (such as peak E
in Fig.1), since at that energy the spectra might be contaminated
with overlapping Auger emissions.   It is instead best to choose
the incident photon energy as low as possible which still gives
appreciable resonance effect.   Some of the previous Ce 3$d$-edge
RPES experiments may have been performed at the photon energy of
maximum resonance, in which case the data have to be interpreted
with caution. This interpretation on the origin of the lineshape
change with the incident photon energy can also explain the dip
structure near E$_F$ shown in Fig.3. Because of the Auger
transition between multiplets, the intensity of the Kondo
resonance peak near E$_F$ will be reduced while some intensities
move to higher binding energy side. Therefore if each spectrum is
normalized to its maximum height before subtraction, as we did in
our case, a dip will show up near the Fermi level with negative
intensity in the difference spectra.

 This phenomenon of overlapping Auger emissions in the Ce
3$d$-edge RPES spectra is not limited to the particular case of
CeFe$_2$. In Fig. 4 and Fig. 5 are shown the Ce 3$d$-edge RPES
data for CeNi$_2$ and CeSi$_2$, respectively, taken at various
incident photon energies.  Again, we see that they both show
changes of spectral shapes depending on the photon energy, similar
to the case of CeFe$_2$.  The two spectra of CeNi$_2$ in Fig.4 are
obtained with the incident photon energy of 882.34eV (E) and
881.54eV (A), which are very close to each other and roughly
corresponds to two maximum points of $3d_{5/2}$ XAS spectrum of
CeNi$_2$ in Fig. 1. But these two 4$f$ spectral weights have
appreciably different shapes in that the spectrum at photon energy
E shows a fairly strong peak around 0.8eV binding energy, which
becomes very weak or absent in the spectrum A.  To understand the
origin of this change of spectral shape, we again took the
difference spectrum of the two and showed it at the bottom of Fig.
4.  We can see in this difference spectrum a broad peak with its
maximum around 0.8eV binding energy, which is exactly the
difference of the photon energy and therefore corresponds to the
constant kinetic energy feature. This tells us that in this case
of CeNi$_2$ as well the origin of this phenomenon is the same as
the case of CeFe$_2$ discussed earlier, and the overlapping Auger
emissions between different multiplets of $3d^94f^2$ intermediate
state contribute to the lineshape change with the incident photon
energy.  The CeSi$_2$ spectra shown in Fig. 5 gives the same
story, although the phenomenon does not look as pronounced as in
the case of CeNi$_2$.

\section{CONCLUSION}

In the newly-developed Ce 3$d$-edge resonance photoemission
technique for the study of bulk electronic structures of Ce
compounds, the spectral shapes were found to depend sensitively on
the incident photon energy.  By analyzing Ce 3$d$-edge RPES
valence band spectra of CeFe$_2$, CeNi$_2$, and CeSi$_2$ compounds
systematically, we showed that this phenomenon originated from the
Auger transition between different multiplet structures of
$\underline{3d}4f^2$ electronic configuration in the intermediate
state of the resonance process.  Although we only showed the data
on CeFe$_2$, CeNi$_2$, and CeSi$_2$ in this paper, we found
similar results for other Ce compounds as well, and expect this
phenomenon to be universal for any Ce compounds. Hence, when the
Ce $f$ spectral weights are obtained with the Ce $3d$-edge RPES
technique, it is very important to avoid these overlapping Auger
emissions contributing to the spectra.  The most logical choice of
the incident photon energy must be that corresponding to the
lowest multiplet of the 3$d$-edge x-ray absorption spectra.

\vspace*{4mm}

\acknowledgements

 One of us(EJCho) acknowledge the support by Korea Research
 Foundation Grant(KRF-2001-015-DP0178). This work is supported by
 the Korean Science and Engineering Foundation (KOSEF) through the
 Center for Strongly Correlated Materials Research (CSCMR) at Seoul
 National University. The research was performed under the support
 of Japan Synchrotron Radiation Research Institute(JASRI).


\newpage

\begin{center}
\LARGE{\bf Figure Captions}
\end{center}

Fig. 1 : x-ray absorption spectra of $3d \rightarrow 4f$
transition region for CeFe$_2$, CeNi$_2$, and CeSi$_2$. The
symbols marked on the spectrum represent photon energies used for
RPES shown in subsequent figures. The bar diagram at the bottom is
the theoretical calculation result for the multiplet structure of
$\underline{3d}f^2$ intermediate state taken from ref. [26].  \\

Fig. 2 : Ce $f$ spectral weights near the Fermi level for CeFe$_2$
obtained with various incident photon energies.  \\

Fig. 3 : Difference spectra between the 4$f$ spectral weights at
various photon energies and that at h$\nu$ = 881.76eV. \\

Fig. 4 : Ce 4$f$ spectral weights near the Fermi level for
CeNi$_2$ obtained at different photon energies.  Shown in the
bottom is the difference spectrum between the two. \\

Fig. 5 : Ce 4$f$ spectral weight near the fermi level for CeSi$_2$
obtained at various photon energies. In the bottom the difference
spectra are shown. \\

\end{document}